\documentclass[11pt]{article}
\usepackage{graphics, feynmp}
\newcommand{\be}{\begin{equation}}
\newcommand{\ee}{\end{equation}}
\newcommand{\bea}{\begin{eqnarray}}
\newcommand{\eea}{\end{eqnarray}}

\begin{document}
\begin{titlepage}
\def\thepage {}        % Kill page numbering

\title{Top-Bottom Color and Weak Doublets Seesaw\thanks{Talk given at the \textit{Thinkshop$^2$-top quark physics for RUN II and beyond}, Fermilab, November 10-12, 2000.}}

\author{
Marko B. Popovic\thanks{e-mail address: 
markopop@buphy.bu.edu},\\
Department of Physics, Boston University, \\
590 Commonwealth Ave., Boston MA  02215}

\date{\today}

\maketitle

\bigskip
\begin{picture}(0,0)(0,0)
\put(295,250){BUHEP-01-2}
\put(295,235){hep-ph/0102027}
\end{picture}
%\vspace{-12pt}
\vspace{24pt}

\begin{abstract}

A brief overview of some of the last decade's developments in the
Top-mode/Topcolor class of models is presented. In addition, a new Topcolor
type scenario, named Top-Bottom Color, is suggested. A seesaw mechanism with new weak doublet quarks has been introduced to lower the previous theoretical
prediction for the large top quark mass; without a seesaw mechanism  and with
a low scale of Nambu - Jona-Lasinio triggering interactions, the top quark
mass was predicted to be \textit{O}($500$GeV). In Top-Bottom Color an effective, composite two Higgs doublet model is obtained where the third generation
isospin splitting is introduced via tilting interactions related to the
broken non-abelian gauge groups (i.e. without strong, triviality-sensitive
$U(1)$ groups). To complete discussion a few notes on the fine
tuning related to the bottom quark mass are appended in an \textit{addendum}. 
A generic problem (\textit{Why does the top Yukawa coupling equal one?}) of
the Top-mode/Topcolor class of models is pointed out as well.
Furthermore, previous work on the third generation seesaw mechanism with weak
doublet quarks and dynamical top mass production in the Topcolor spirit is discussed.

\pagestyle{empty}
\end{abstract}
\end{titlepage}

%%%%%%%%%%%%%%%%%%%%%%%%%%%%%%%%%%%

%%%%%%%%%%%%%%%%%%%%%%%%%%%%%%%%%%%

\section{Introduction: Top-mode and Topcolor models in the last decade - a brief (and incomplete) overview}
\label{sec:intro}
\setcounter{equation}{0}

Taking the four-dimensionality of our physical space for granted necessarily
yields to the famous hierarchy problem in the standard (non-supersymmetric)
gauge theories. The Higgs scalar is quadratically unstable against radiative
corrections, and one naturally expects its mass to be of the same (or similar) order as the highest scale of the theory (presumably the Plack scale $\sim 10^{19}$GeV) \cite{unnatural}.  

Models with strong dynamics, like Technicolor (TC) \cite{tc}, address the
hierarchy problem through the slow, logarithmic running of gauge couplings of
non-abelian gauge groups that, after a huge interval of energy scales,
finally become strong enough to produce substantial, ``interesting", effects
on our low-energy world. The main effect is the presence of \textit{condensates}
- strongly interacting \textit{new} fermions bind to form condensates that
break the global symmetries giving Nambu-Goldston bosons with appropriate
quantum numbers (to be \textit{eaten} by $Z$ and $W$); these theories do not posess a \textit{fundamental} Higgs field floating at the highest energy scales.

Moreover, the fermion masses can be explained in an extended framework of this dynamical scheme - in the spirit of Extended Technicolor (ETC)
\cite{etc}. The standard model (SM) fermions obtain their masses through
Yukawa type couplings with now `composite scalar fields', implying the
presence of four-fermion interactions (from broken ETC, i.e. extended TC gauge group
structure) at higher energies. Although in many ways attractive, this idea
carries a number of serious challenges and drawbacks - particularly in the
simultaneous dynamical breaking of EW symmetry and the creation of mass for third generation quarks (for a review see \cite{course}). 

Attempts to introduce a composite scalar Higgs made of third generation SM
fermions are expressed in the class of Top-mode models \cite{top-mode,
  top-mode1}. The Top-mode scheme \cite{top-mode1} suggests the existence of 4-fermion interactions at some high-energy scale $\Lambda$, i.e.
\be
\mathit{L_{4-fermion}}={g^2 \over m_0^2} \, \left({\overline{\Psi}}_{Li}^a t_{Ra} \right) \, \left(\overline{t}_{R}^b \Psi_{Lb}^i \right) \; .
\ee
where $\Psi_L=(t, b)_L$ and where the index $i$ ($a,b$) labels $SU(2)_W$
($SU(3)_{QCD}$) elements in the fundamental representation. The mass $m_0$ is of
the same (or similar) order as the scale $\Lambda$. The coupling $g^2$ is
given \textit{ad hoc} (without dynamical explanation) and it is assumed large. The interaction lagrangian may be rewritten \cite{top-mode1} (without changing the equation of motion) as 
\be
\mathit{L_{4-fermion}}= -g \left({\overline{\Psi}}_{L} t_{R} \Phi + h.c. \right) - M_0^2 \Phi^{\dagger} \Phi 
\ee
where the auxiliary \textit{static} scalar field $\Phi$ is a \textit{would-be}
Higgs scalar field at low-energy. The field $\Phi$ gets a gauge invariant
kinetic term, a positive contribution to the quadratic term and a negative
$\Phi^4$ term through the fermion loops in the block-spin renormalization
scheme, while sliding to lower energy scales. The predicted top quark mass,
in the ``full SM" analysis \cite{top-mode1} with a fixed value of the EW VEV,
$v_{EW}\approx250$GeV, was found to be $O$($500$GeV) for a triggering
interaction at  $\Lambda\sim10^4$GeV (with very large, though still benign fine
tuning\footnote{Expressed as a small relative
  deviation of coupling $g^2$ from the critical coupling $g_c^2$ (needed to
  trigger condensation), as predicted for example by the gap equation in the
  Nambu - Jona-Lasinio (NJL) approximation \cite{NJL}.} ) and $O$($200$GeV)
(but still larger than the physical top quark mass) for $\Lambda$ of the
order of the Planck scale\footnote{This author is not fully convinced  of the
  correctness of the results obtained in the ``full SM" analysis
  \cite{top-mode1} because of the specific handling of the composite scalar
  loops. We agree with reference \cite{top-mode1} that scalar loops in
  principle should be considered (for calculation of, say, the $\beta$
  function) - but only when the effective, scale dependent, mass
  $m_{H{eff}}(\Lambda')$ is smaller than the renormalization scale
  $\Lambda'$, and not all the way up to the compositeness  scale
  $\Lambda$. Therefore, we would rather predict $O$($200$GeV) for the top
  quark mass when $\Lambda\sim10$TeV (instead of the cited $O$($500$GeV)),
  which follows from the Pagels-Stokar relationship \cite{Pagels-Stokar} with
  a fixed value of the top composite scalar VEV, $v_t=v_{EW}\approx250$GeV. We plan to address these issues in more detail elsewhere.} (with tremendous fine tuning). We illustrate this result in Fig. 1.

\begin{figure}
\begin{center}
\rotatebox{0}{\scalebox{.5}{\includegraphics{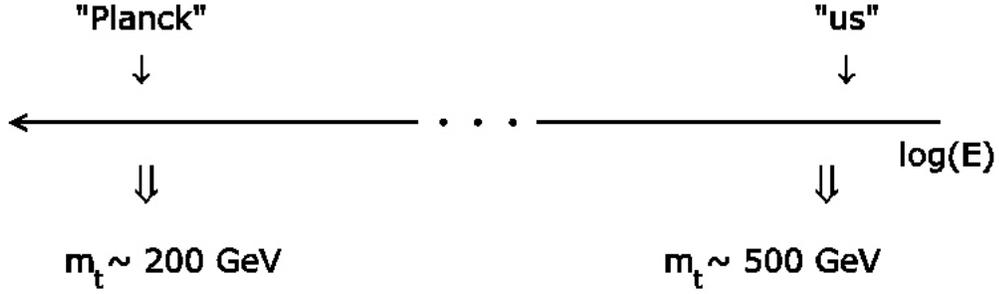}}}
\end{center}
\caption[lepton]{Illustration of the predicted \cite{top-mode1} mass of the top quark with respect to the NJL triggering scale.}
\label{talk1}
\end{figure}

A dynamical basis for the \textit{ad hoc} interaction (1.1) is
introduced through the presence of an additional strong $SU(3)$ interaction
called Topcolor \cite{topc}. The assumed gauge symmetry breaking pattern
$SU(3)_A \otimes SU(3)_B \rightarrow SU(3)_{QCD}$ produces an octet of heavy
colorons \cite{coloron} that may act as an NJL binding force for the fermions
transforming under the stronger ``initial" $SU(3)$ gauge group. The recipe is
simple - at a scale somewhat below the coloron mass, the coloron exchange
diagrams are approximated by four-fermion interactions; fiertzing
interactions that couple the lefthanded and righthanded currents in the large
$N_c$ limit ($N_c$ is the number of colors, in our case $N_c=3$) gives the
interaction term (1.1). 

The Topcolor dynamics was used, combined with TC/ETC dynamics, in third
generation specific\footnote{Where only third generation quarks feel the
  stronger ``initial" $SU(3)$ gauge group.} \cite{TC2, TC22} Topcolor
Assisted Technicolor models (TC$^2$) as well as in the
universal\footnote{Where all SM quarks feel the stronger ``initial" $SU(3)$
  gauge group.} \cite{pap1} TC$^2$ models in order to create a heavy top
quark mass (with top composite scalar VEV, $v_t$, usually expected to be of order $v_{EW}/3$). 

\begin{figure}
\begin{center}
\rotatebox{0}{\scalebox{.5}{\includegraphics{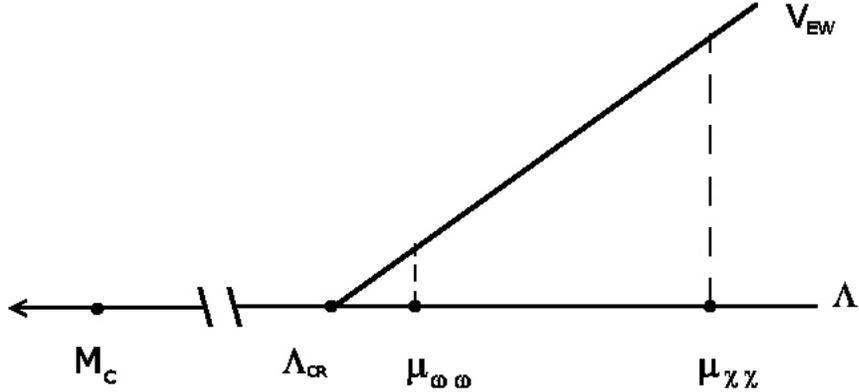}}}
\end{center}
\caption[lepton]{The schematic illustration of the mass renormalization curves in the weak-singlet seesaw Topcolor models \cite{topseesaw1, topseesaw2}.}
\label{talk2}
\end{figure}

The general motivation for the weak-singlet seesaw models
\cite{topseesaw1,topseesaw2} involving third generation quarks and embedded
in a Topcolor scheme was to ``lower" the earlier prediction for
the top quark mass when $v_t=v_{EW}$ (while keeping the scale of Topcolor low).
The seesaw matrix with new weak-singlet quarks $\chi_L$ and $\chi_R$ was introduced in the following form
\bea
( \; \overline{t}_L \;\;\; \overline{\chi}_L \; ) \left( \begin{array}{cc} 0 & m_{t\chi} \\ \mu_{t\chi} & \mu_{\chi\chi} \end{array} \right) \left( \begin{array}{c} t_R \\ \chi_R \end{array} \right) .
\eea
The mass term $m_{t\chi}$ is directly related to the Topcolor dynamics
(i.e. $\chi_R$ and $\Psi_L=(t,b)_L$ feel the stronger ``initial" $SU(3)$
gauge group) while for the purpose of our discussion $\mu_{t\chi}$ and
$\mu_{\chi\chi}$ may be thought to be the bare masses (though only
$\mu_{t\chi}$ may be a fundamental bare mass, i.e. $\chi_L$ and $t_R$ feel
the weaker ``initial'' $SU(3)$ gauge group). Sliding down from
the triggering scale ($\Lambda<M_c$, where $M_c$ is the coloron mass) the fermion loop induced renormalization dynamics is
stopped at the scale  $\sqrt{\mu_{t\chi}^2 + \mu_{\chi\chi}^2}
\approx \mu_{\chi\chi}$ (assuming $\mu_{\chi\chi} \gg \mu_{t\chi}$). The
fermion loop calculation is then used in obtaining the conditions for proper
EW symmetry breaking and the mass spectrum of scalar states
\cite{topseesaw2}\footnote{This author is not fully convinced
  that the block-spin renormalization approach with
  the fermion loop contribution alone is justified. It seems that $\mu_{\chi\chi}$ is large
  enough so that the $\Phi^4$ term contribution to $m_{H{eff}}(\Lambda')$ may
  be neglected and that the scalar loop contributions should then be considered at
  low enough energy scales (larger than $\mu_{\chi\chi}$) - in the framework
  with a hard momentum cut-off.}. In addition, taking $m_{t\chi}\approx
600$GeV as a working premise demands \cite{topseesaw2} $m_{\chi\chi}$ be at
least $5$TeV as implied by the limits on the $T$ parameter. It was
suggested \cite{topseesaw2} that a similar mechanism may be used for the generation of bottom quark mass as well - one may add the
new weak-singlet quarks $\omega_L$ and  $\omega_R$ and two new ``bare" mass
terms $\mu_{\omega\omega}$ and $\mu_{b\omega}$ (with
$\sqrt{\mu_{\omega\omega}^2 + \mu_{b\omega}^2} \approx \mu_{\omega\omega}$)
to the model structure. With the same NJL triggering interactions in the top and
bottom sector, the top and bottom mass renormalization curves share the same
path up to the mass $\mu_{\omega\omega}$ where dynamics in the bottom sector
stops. This behavior is illustrated in Fig. 2. The scale $\Lambda_{cr}$ represents
the critical scale at which (in the scheme with a hard momentum cut-off, as in
\cite{top-mode1, topseesaw1, topseesaw2}), the dynamical fermion mass starts
to grow rapidly.  The window between
$\Lambda_{cr}$ and $\mu_{\chi\chi}$, in this approximation, is extremely narrow
($O$($200$GeV) for $\mu_{\chi\chi}\approx5$TeV). For our purposes the
renormalization mass curve may be thought to be linear\footnote{Clearly, the dynamical mass may be then related to $\tan{\beta}={m_{t\chi} \over m_{b\omega}} = {{\Lambda_{cr} - \mu_{\chi\chi}} \over {\Lambda_{cr} -\mu_{\omega\omega}}}$.}.

\begin{figure}
\begin{center}
\rotatebox{0}{\scalebox{.5}{\includegraphics{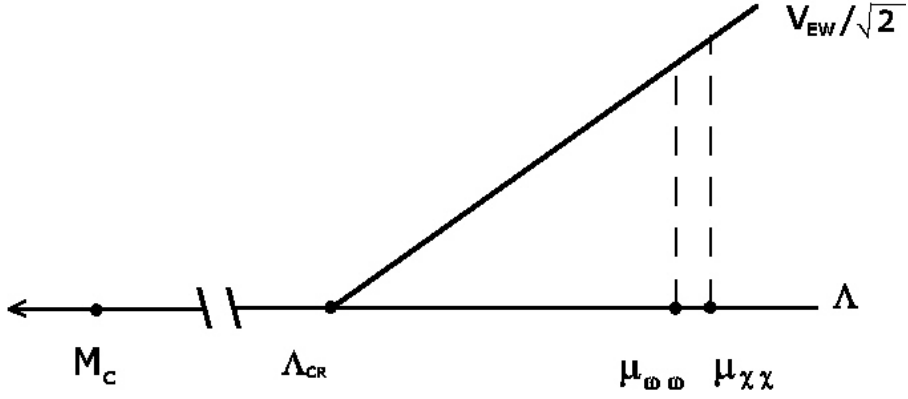}}}
\end{center}
\caption[lepton]{The schematic illustration of the mass renormalization
  curves in the ``two-doublet" model \cite{2see}.}
\label{talk3}
\end{figure}

Another attempt, reflected in a two-doublet model \cite{2see}, to
introduce the bottom quark mass generation in the Topcolor scheme with
a weak-singlet seesaw, in a manner similar to the one described above, assumes
$\tan{\beta}=v_t/v_b=1$ (where $v_t,v_b=v_{EW}/\sqrt{2}$ are the VEVs in the
top and bottom sector). This clearly implies that
$\mu_{\chi\chi}=\mu_{\omega\omega}$ and $m_{t\chi}=m_{b\omega}$. This
situation is illustrated in Fig. 3. Therefore
$m_t/m_b=\mu_{t\chi}/\mu_{b\omega}$. The most stringent constraint comes from
the parameter $R_b$ and the limit on ``bare'' masses,
$\mu_{\chi\chi}=\mu_{\omega\omega}$, is now pushed up to $12-15$TeV \cite{2see} (this
implies an even smaller window between $\Lambda_{cr}$ and the heavy ``bare''
mass than in the previous case - by roughly a factor $3\sqrt{2}$)\footnote{As
  the heavy masses are very large here (setting the scale of  mass generation
  dynamics) the effect of the $\Phi^4$ term contribution to the effective
  mass may be safely neglected and the scalar loops, we believe, should
  definitely be considered.}.

\section{Top-Bottom Color and Weak-Doublets Seesaw}
\label{sec:T-B Color}
\setcounter{equation}{0}

We consider\footnote{We are not first to introduce this extended third generation seesaw
  structure with weak-doublet quarks in the Topcolor type of models. As we learned at some point after we
  presented this material at the \textit{Thinkshop$^2$}, an identical
  low-energy structure with weak-doublet quarks has been introduced in the spirit
  of a third generation specific TC$^2$ strong dynamical scheme (with
  identical extended Topcolor gauge group sector but with $v_t\approx v_{EW}$) in the work
  of the He, Tait and Yuan, \cite{topflavor}. We thank H.-J. He for informing
  us about his past work. The main differences between our strong dynamics
  scenario (Top-Bottom Color) and the one in reference \cite{topflavor} are:
  $1)$ instead of using a strong $U(1)$ gauge group - that necessarily has to be
  very strong to avoid a fine tuning problem (and therefore, we believe, it
  is not easily accommodated in a natural dynamical scheme) - we introduced
  the top-bottom mass splitting through the effect of an additional strong,
  asymptotically free, non-abelian $SU(3)$ gauge group, and $2)$ we suggest the  bottom quark mass generation via the same Topcolor type of mechanism while reference \cite{topflavor} suggests that the bottom quark mass must be generated by different mechanism.} an extended third
  generation quark sector with new weak-doublet quarks transforming under the
  SM gauge group ($SU(3)_{QCD} \otimes SU(2)_W \otimes U(1)_Y$) as
\bea
{\Psi_1}_L=\left( \begin{array}{c} t_L \\ b_L \end{array} \right)\, , \, (3,2,1/6); \; &
{\Psi_2}_L=\left( \begin{array}{c} T_L \\ B_L \end{array} \right)\, , \, (3,2,1/6); \nonumber \\ 
\Psi_R=\left( \begin{array}{c} T_R \\ B_R \end{array} \right)\, , \,  (3,2,1/6); \; & t_R \, , \, (3,1,2/3); \; \; \; b_R \, , \, (3,1,-1/3).
\eea
The mass matrices in the top and bottom sectors are 
\bea
( \; \overline{t}_L \;\;\; \overline{T}_L \; ) \left( \begin{array}{cc} m_1 & m_p \\ 0 & m_q \end{array} \right) \left( \begin{array}{c} t_R \\ T_R \end{array} \right) \; \\
( \; \overline{b}_L \;\;\; \overline{B}_L \; ) \left( \begin{array}{cc} m_2 & m_p \\ 0 & m_q \end{array} \right) \left( \begin{array}{c} b_R \\ B_R \end{array} \right) .
\eea
At this point we imagine $m_p$ and $m_q$ to be the ``bare'', weak
  doublet-doublet, mass terms and $m_1$ and $m_2$ to be the (Topcolor)
  dynamical, weak singlet-doublet, mass terms. In addition we assume
  $m_p>m_q>m_1>m_2$. Performing separately the rotations of the lefthanded
  (by angle $\varphi_q^L$) and righthanded (by angle $\varphi_q$) fermions the above mass matrices are diagonalized and we obtain the physical light and heavy masses, i.e.
\be
m_t \approx {{m_1 m_q} \over \sqrt{m_p^2 + m_q^2}} \;\; ; \;\; m_b \approx {{m_2 m_q} \over \sqrt{m_p^2 + m_q^2}} \;\; ; \;\; M_T \approx M_B \approx M = \sqrt{m_p^2 + m_q^2}
\ee

The EW precision measurements yield a $3\sigma$ limit on the righthanded bottom
mixing - we find \cite{tbcolor} $\sin^2{\varphi_b}<0.0052$ while known
physical top and bottom quark masses yield a more stingent\footnote{The
  stringent limit \cite{tbcolor} is obtained by assuming the Topcolor value $m_1=600$GeV, giving $\sin^2{\varphi_b}<10^{-5}$.} consistency limit \cite{tbcolor}, i.e. $\sin^2{\varphi_b}<0.0006$.

\begin{figure}
\begin{center}
\rotatebox{0}{\scalebox{.5}{\includegraphics{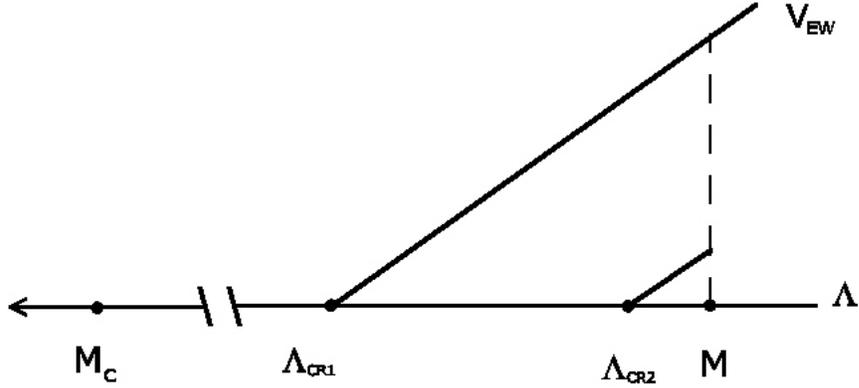}}}
\end{center}
\caption[lepton]{The schematic illustration of the mass renormalization
  curves in the Top-Bottom Color scenario.}
\label{talk4}
\end{figure}

Clearly, both the top and bottom sectors have the same infrared cut-off ($\sim
M$) as illustrated in Fig. 4. Therefore, one needs different NJL interactions in the top and bottom sectors in order to introduce the isospin mass splitting. As shown in Fig. 4 two different ``critical" energy scales, $\Lambda_{cr1}$ and $\Lambda_{cr2}$ are present, corresponding to the top and bottom mass renormalization curves respectively. 

\begin{figure}
\begin{center}
\rotatebox{0}{\scalebox{.5}{\includegraphics{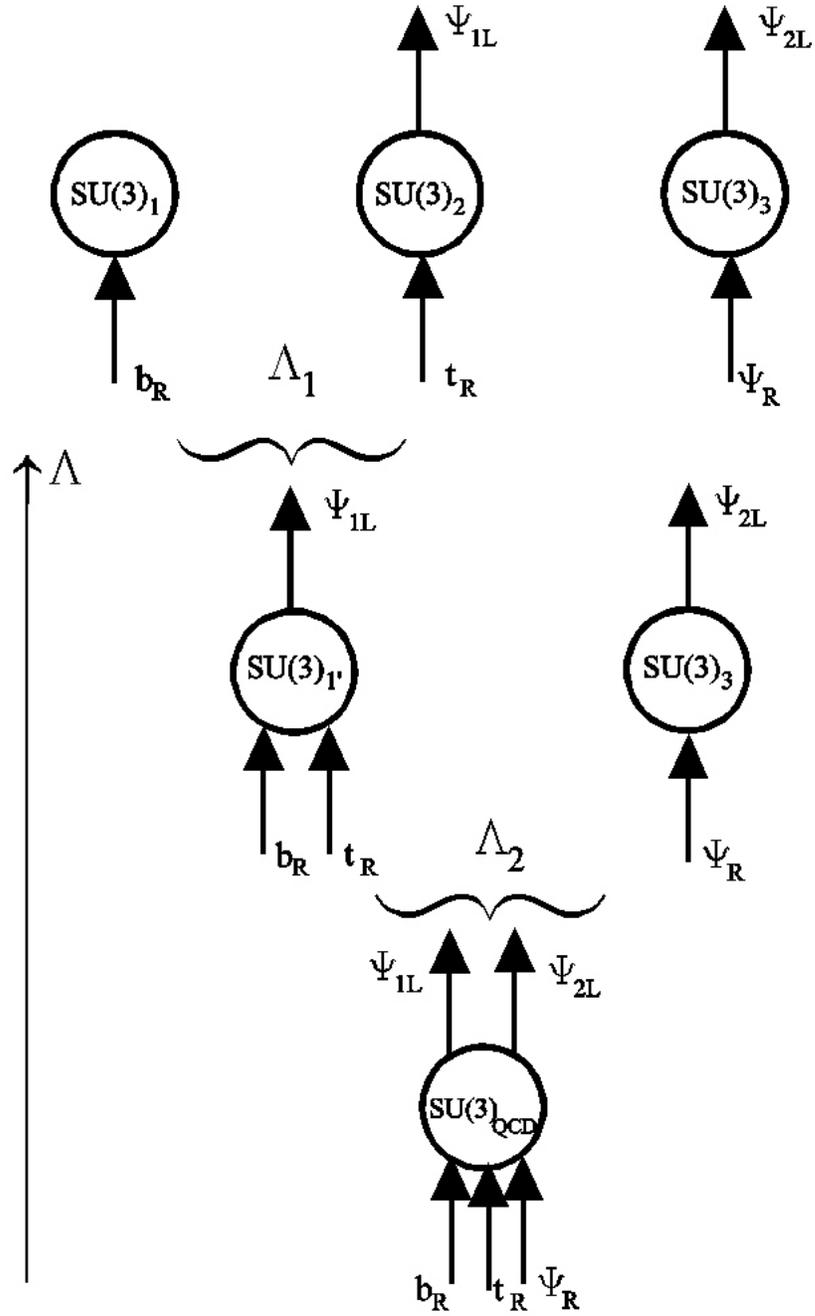}}}
\end{center}
\caption[lepton]{The schematic illustration of the cascade of symmetry
  breakings and fermion charge assignments under the strong
  $SU(3)$ gauge groups in the Top-Bottom Color scenario.}
\label{Moose2}
\end{figure}

The third generation isospin splitting is introduced \cite{tbcolor} via a strong $SU(3)$ gauge group in addition to the Topcolor structure (therefore the name - Top-Bottom Color). The third generation quark sector introduced in the low energy setup above transforms under this extended gauge structure as 
\bea
{\Psi_1}_L=\left( \begin{array}{c} t_L \\ b_L \end{array} \right)\, , \, (1,3,1,2,1/6); \;\;\;
{\Psi_2}_L=\left( \begin{array}{c} T_L \\ B_L \end{array} \right)\, , \, (1,1,3,2,1/6); \nonumber \\ 
\Psi_R=\left( \begin{array}{c} T_R \\ B_R \end{array} \right)\,  , \,  (1,1,3,2,1/6); \; \; t_R \, , \, (1,3,1,1,2/3); \; \; b_R \, , \, (3,1,1,1,-1/3);
\eea
where instead of gauge assignments under $SU(3)_{QCD}$, as in equation (2.1), we introduce the assignments under the strong $SU(3)$ gauge ``trio". The schematic illustration of the cascade of symmetry breakings and fermion charge assignments under the strong $SU(3)$ gauge groups is shown in Fig. 5.

Defining the gauge couplings of this strong ``trio" as $g_1,g_2$ and $g_3$ we find
the tilting interaction below the higher symmetry breaking scale $\Lambda_1$
(corresponding to $SU(3)_1 \otimes SU(3)_2 \rightarrow SU(3)_{1'}$ and giving
an octet of \textit{precolorons}, i.e. heavier colorons) to be proportional to 
\be
{g_2^2 \over {M_{pc}^2 (g_1^2 + g_2^2)}} \left[ g_2^2 \left(
    {\overline{\Psi}}_{1L} t_R \right) \left( {\overline{t}}_R \Psi_{1L}
  \right) - g_1^2 \left( {\overline{\Psi}}_{1L} b_R \right) \left(
    {\overline{b}}_R \Psi_{1L} \right) \right] \; .
\ee
where $M_{pc}$ is the precoloron mass.
However, we assume that these interactions are not strong enough to trigger dynamical condensation. Nonetheless, they represent a crucial tilting needed for isospin splitting of the third generation quarks. 

The coupling $g_{1'}$ (the $SU(3)_{1'}$ gauge coupling) runs enough below
$\Lambda_1$ so that the NJL interactions triggered at the lower scale
$\Lambda_2$ (corresponding to $SU(3)_{1'} \otimes SU(3)_3 \rightarrow
SU(3)_{QCD}$ breaking) are strong enough to produce condensates in both the
top and bottom sectors\footnote{A more detailed description of this dynamics
  may be found in reference \cite{tbcolor}, illustrated with an anomaly free, example model structure.} in the standard manner of Topcolor models. Therefore, we obtain an effective composite two Higgs doublet model with $\tan{\beta}=m_t/m_b$.

\textit{Addendum} The generation of bottom (in addition to top) quark mass in
Topcolor models certainly introduce a new amount of fine tuning in the
theory. Whether this represents a drawback for model structure or the natural
consequence of an additional parameter in the theory is yet to be
understood. Certainly, the fine tuning is much smaller than with a triggering
scale of order the Planck mass (and the creation of the top quark mass
alone). Anyhow, we find that as a lesser problem than the missing explanation
of \textit{why the top Yukawa coupling equals one}. Stated differently, the
successful seesaw model, for example, definitely needs to explain in a
natural manner why the ``bare" masses relate \textit{conspiratorially} when
the dynamical mass has the  fixed value of say  $O$($600$GeV). 
The possibility that the bottom mass (in our case dynamical mass $m_2 \geq 12$GeV) may have a different origin - either through ETC contributions, Yukawa couplings with composite \textit{top} Higgs scalar (from the higher dimensional effective operators) or instanton contributions\footnote{We thank C. T. Hill for pointing us to the importance of this possibility.} - should be carefully considered as well. 

\textit{I thank R. S. Chivukula and E. H. Simmons for valuable and
  motivating discussions and K. R. Lynch for helpful comments on the manuscript. This work was supported in part 
by the National Science Foundation under grant PHY-9501249, and by the
Department of Energy under grant DE-FG02-91ER40676. }


\begin{thebibliography}{9}

\bibitem{unnatural} K. G. Wilson, unpublished; quoted in L. Susskind, Phys. Rev. {\bf D20}, 2619 (1979); G. 't Hooft, in \textit{Recent Developments in Gauge Theories}, edited by G. 't Hooft at al. , \textit{Plenum} New York, (1980).
\bibitem{tc} S. Weinberg, Phys. Rev. D {\bf 13}, 974 (1976); {\bf 19}, 1277
  (1979); L. Susskind, {\it ibid.} {\bf 20}, 2619 (1979).
\bibitem{etc} S. Dimopoulos and L. Susskind, Nucl. Phys. {\bf B155}, 237
  (1979); E. Eichten and K. Lane, Phys. Lett. {\bf 90B}, 125 (1980).
\bibitem{course} R. S. Chivukula, Lectures presented at 1997 Les Houches Summer School, [hep-ph/9803219].
\bibitem{top-mode} V. A. Miransky, M. Tabanashi and K. Yamawaki,
  Phys. Lett. B {\bf 221}, 177 (1989); Mod. Phys. Lett. A {\bf 4}, 1043 (1989);
  Y. Nambu, Chicago report EFI 89-08 (1989); W. J. Marciano,
  Phys. Rev. Lett. {\bf 62}, 2793 (1989); Phys. Rev. D {\bf 41}, 219 (1990).
\bibitem{top-mode1}  W. A. Bardeen, C. T. Hill and M. Lindner, Phys. Rev. {\bf  D 41}, 1647 (1990).
\bibitem{NJL} Y. Nambu and G. Jona-Lasinio, Phys. Rev. {\bf 122}, 345 (1961);
  {\bf 124}, 246 (1961).
\bibitem{Pagels-Stokar} H. Pagels and S. Stokar, Phys. Rev. {\bf D20}, 2947 (1979).
\bibitem{topc} C. T. Hill, Phys. Lett. B {\bf 266}, 419 (1991).
\bibitem{coloron} R. S. Chivukula, A. G. Cohen and E. H. Simmons, Phys. Lett. B {\bf 380}, 92 (1996).
\bibitem{TC2} C. T. Hill, Phys. Lett. B {\bf 345}, 483 (1995). 
\bibitem{TC22} K. Lane and E. Eichten, Phys. Lett. B {\bf 352}, 382 (1995); K. D. Lane, Phys. Rev. D {\bf 54}, 2204 (1996); G. Buchalla {\it et al.}, Phys. Rev. D {\bf 53}, 5185 (1996).
\bibitem{pap1} M. B. Popovic and E. H. Simmons, Phys. Rev. {\bf D58}, 095007 (1998) [hep-ph/9806287].
\bibitem{topseesaw1} B. A. Dobrescu and C. T. Hill, Phys. Rev. Lett. {\bf 81},
  2634 (1998) [hep-ph/9712319]. 
\bibitem{topseesaw2} R. S. Chivukula, B. A. Dobrescu, H. Georgi and C. T. Hill, Phys. Rev. {\bf D59}, 075003 (1999) [hep-ph/9809470].
\bibitem{2see} H. Collins, A. Grant and H. Georgi, Phys. Rev. {\bf D61},
  055002 (2000) [hep-ph/9908330].
\bibitem{topflavor} H.-J. He, T. M. P. Tait and C.-P. Yuan, Phys. Rev. {\bf D62}, 011702 (2000) [hep-ph/9911266].
\bibitem{tbcolor} M. B. Popovic, (2000) [hep-ph/0101123].

\end{thebibliography}
\end{document}